\documentstyle[psfig,epsf]{mn}
\newcommand{\simgt}{\lower.5ex\hbox{$\; \buildrel > \over \sim \;$}}
\newcommand{\simlt}{\lower.5ex\hbox{$\; \buildrel < \over \sim \;$}}

\def\impc{ h {\rm Mpc^{-1}}}
\def\r{{\bf {r}}}
\def\k{{\bf {k}}}
\def\mpc{ h^{-1} {\rm Mpc}}

\def\kms{\,{\rm {km\, s^{-1}}}}
\def\msun{{M_\odot}}
\newcommand{\citet}[1] {\cite{#1}}
\newcommand{\citep}[1] {(\cite{#1})}
\begin{document}

\title[An analytical model for the non-linear redshift-space power spectrum]{An analytical model for the non-linear redshift-space power spectrum}
\author[Kang, Jing, Mo \& Borner]{Xi Kang$^{1}$, Y.P. Jing$^{1,2}$, H. J. Mo$^{2}$, and  G. B\"orner$^{2}$\\
$^{1}$Shanghai Astronomical Observatory, the Partner Group of MPI f\"ur
Astrophysik, Nandan Road 80, Shanghai 200030, China\\
$^{2}$ Max-Planck Institut f\"ur Astrophysik, Karl-Schwarzschild-Strasse
1, 85748 Garching, Germany\\
\smallskip
Email: kangx@center.shao.ac.cn, ypjing@center.shao.ac.cn,
hom@mpa-garching.mpg.de, grb@mpa-garching.mpg.de}

\maketitle

\begin{abstract}
We use N-body simulations to test the predictions of the redshift
distortion in the power spectrum given by the halo model in which the
clustering of dark matter particles is considered as a result both of the
clustering of dark halos in space and of the distribution of dark matter
particles in individual dark halo. The predicted redshift distortion
depends sensitively on several model parameters in a way different
from the real-space power spectrum. An accurate model of the redshift
distortion can be constructed if the following properties of the halo
population are modelled accurately: the mass function of dark halos,
the velocity dispersion among dark halos, and the non-linear nature of
halo bias on small scales. The model can be readily applied to
interpreting the clustering properties and velocity dispersion of
different populations of galaxies once a cluster-weighted bias (or
equivalently an halo occupation number model) is specified for the
galaxies. Some non-trivial bias features observed from redshift surveys
of optical galaxies and of IRAS galaxies relative to the standard
low-density cold dark matter model can be easily explained in the
cluster weighted bias model. The halo model further indicates that a
linear bias can be a good approximation only on for $k \leq 0.1 hMpc^{-1}$.
\end{abstract}

\section{Introduction}
The power spectrum of the galaxy spatial distribution is an important
statistic for describing inhomogeneities in the Universe. The
spatial distribution of galaxies observed with a redshift survey is
distorted by the peculiar motions of galaxies, and a
statistically isotropic distribution (e.g. the power spectrum or the
correlation function) in real space becomes anisotropic in
redshift space (Geller \& Peebles 1973; Davis \& Peebles 1983; Bean et
al. 1983; Kaiser 1987). On the other hand, when measuring
clustering for high redshift objects in redshift space, choosing the
wrong cosmological model can lead to an additional anisotropy in the
redshift space distribution (Matsubara \& Suto 1996; Ballinger,
Peacock \& Heavens 1996).
The theory for the redshift distortions caused by an assumed world
model and by the large scale linear motions is now well
established. The transform of the clustering pattern from the true
world model to an assumed model is just a simple mapping in the
coordinates. On large scales and for a linear bias, the redshift power
spectrum $P^{S}_{l}(k, \mu)$ can be derived (Kaiser 1987)
\begin{equation}
P^{S}_{l}(k, \mu) = P^{R}_{l}(k)\, [1+\beta \mu^{2}]^{2}
\label{Pk}
\end{equation}
where $\mu$ is the cosine of the angle between the line-of-sight
and the $\bf k$ \rm vector, $\beta=\Omega^{0.6}/b$, $\Omega$ the
density parameter, $b$ the linear bias, and $P^{R}_{l}(k)$ the
linear power spectrum of dark matter in real space. The hope has
 been to measure the dynamical quantity $\beta$ and the cosmological 
parameters through studying the anisotropy of galaxies or galaxy cluster
on large enough scales. However, it has been
shown that the virialized motions within rich clusters (the
Finger-of-God effect) are so prominent that the clustering pattern on 
large scales of wave number $k\sim 0.1\impc$, is significantly affected
(Cole, Fisher \& Weinberg 1994; Suto et al. 1999). The effect of
non-linear motions on the redshift distortion must be
properly modelled in order to measure the dynamical quantity and
the cosmological parameters from the redshift distortion of
extragalactic objects. Recent results (Peacock et al. 2001)
from the redshift distortion of the 2dF galaxies further support
this point, as even with the largest redshift survey available
today there exists a tight degeneracy in the determinations of the
parameters $\beta$ and the pairwise velocity $\sigma_v$
(reflecting the non-linear motion).  As pointed out recently by
Jing \& B\"orner (2001, hereafter JB2001), the existing analytical
model for the nonlinear velocity distortion, e.g. the exponential
distribution of the relative velocity in coordinate space
(Davis \& Peebles 1983) or the Lorentz damping function in 
k-space (Cole, Fisher \& Weinberg 1995; Peacock \& Dodds 1994), is at best an
approximation for some scales. Although JB2001 have made an
extensive study of the redshift distortion for the dark matter
and for two specific biased tracers in three typical cosmological
models based on high-resolution simulations, it is unknown how to
generalize their results to cosmological models and/or to biased
tracers different from those studied in their work without running
new simulations. In this paper, we present an analytical model
for the redshift distortion based on the halo model, and we will test
the accuracy of the analytic model with the results of JB2001.

  A key concept in the standard hierarchical scenario of structure
 formation is the formation of dark matter halos, which are virialized
 systems of dark matter particles formed through non-linear
 gravitational collapse in the cosmic density field. Since the
 formation of dark halos involves only gravitational physics, accurate
 analytic models are now available for many properties of the halo
 population, such as the mass function (Press \& Schechter 1974; 
Lee \& Shandarin 1998; Sheth \& Tormen 1999; Sheth, Mo \& Tormen 2001),
 clustering properties, (Mo \& White 1996; Mo, Jing \& White 1997;
 Jing 1998, 1999; Sheth \& Tormen 1999; Sheth, Mo \& Tormen 2001;
 Hamana et al. 2001), and density profiles (Navarro, Frenk \& White
 1996, 1997; Moore et al. 1999; Jing \& Suto 2000; Klypin et
 al. 2001).  Such models are very useful in understanding the
 clustering properties of matter in the universe, as well as in
 understanding the bias of the galaxy distribution relative to the
 underlying density field. Indeed, since in the hierarchical cosmogony
 all masses in the universe are partitioned in dark halos, the halo
 clustering properties , the halo mass function and the halo density 
profiles are  sufficient for the construction of a clustering model
 for the dark matter in the universe (Scherrer \& Bertschinger 1991; 
Sheth \& Jain 1997; Ma \& Fry, 2000a,b; Seljak 2000; Cooray, Hu 
\&  Miralda-Escude 2000; Cooray 2001). Furthermore, since galaxies 
are assumed to form through gas cooling and condensation in dark 
halos, a halo  model of galaxy clustering can also be constructed 
by combining the  clustering properties of the halo population with an assumption of  how galaxies populate dark halos (Jing, Mo \& B\"orner 1998; Peacock
 \& Smith 2000; Seljak 2000; Scocimarro \& Sheth 2001; Scoccimarro et
 al. 2001; Berlind \& Weinberg 2001).

  The halo model also provides a useful way to understand the redshift
 distortion in galaxy clustering (Seljak 2001; White 2001). In this
 model, the enhancement of the redshift-space power spectrum (relative
 to that in real space) on large scales is assumed to arise in the
 halo-halo correlation, while the smearing (Finger-of-God) effect on
 small scales is attributed to the velocity dispersions in dark
 halos. In this paper we show, however, that the redshift distortion
 depends on several important effects which were not included in early
 modelling.  Using results derived from high-resolution N-body
 simulations we show  that an accurate model for the redshift
 distortion dependents not only on the mass function of dark
 halos, the dark halo clustering and the velocity dispersion among
 dark halos, but also on the nonlinear motion of halos at intermediate
 scales.

\section{Power spectrum in real space}
In the halo model, the mass density field in the Universe is a
superposition of halos distributed in space. The mass function
$\phi(M)\,dM$ of halos (which is the mean density of halos with
mass in the range of [$M$,$M+dM$]) and their spatial distributions
$n(M,\r)$\,$dM$ (which is the number density of halos with
mass in the range of [$M$,$M+dM$] at position \r) can be
described by the extended Press-Schechter formalism (Press \&
Schechter 1974; Bond et al. 1991, Bower 1991, Mo \& White 1996) or
its empirical fitting formula from N-body simulations (Jing 1998, 
1999; Sheth \& Torman 1999). Assuming the density
profile of halos to be $\rho_{\alpha}(M,\r)$ for mass $M$, the
density in the Universe can be expressed as
\begin{equation}
\rho(\r)=\int n(M,\r_1) \rho_{\alpha}(M,|\r-\r_1|) dM d^{3}\r_1\,.
\end{equation}
Following Peebles (1980) and Scherrer \& Bertschinger (1991), we can divide the 4-dimensional space of $\r$
and $M$ into infinitesimal volume elements so that the occupation
number $n_{i}$ of halos within any of the volume elements is either $0$ or $1$.
Then the density field can be written as,
\begin{equation}
\rho(\r)=\sum_{i} n_i \rho_{\alpha}(M_i,|\r-\r_i|\rm) \,.
\end{equation}
\label{eq3}
Using the following relations
\begin{eqnarray}
\langle n_i \rangle=\langle n_i^{2} \rangle=\phi(M_i)
dM_i\,d^{3}\r_i\,, \nonumber\\
\langle n_i n_j \rangle_{i\neq j}=\phi(M_{i})
\phi(M_{j})[1+\xi(r_{ij})] d^{3}\r_id^{3}\r_jdM_{i}dM_{j},
\end{eqnarray}
\label{eq4}
where $r_{ij}=|\r_i-\r_j|$, we can derive the two-point correlation
function $\xi(r)$ for the density field which is a sum of two terms,
\begin{equation}
\xi(r)=\xi_{1h}(r)+\xi_{2h}(r)\,
\end{equation}
where the first term on the $rhs$ is the so-called one-halo term $\xi_{1h}(r)$
which is contributed by the internal halo structure (the density
profile) and the second term is the two-halo term $\xi_{2h}(r)$ which
is contributed by the halo-halo clustering (Ma \& Fry 2000b):
\begin{eqnarray}
\bar{\rho}^{2} \xi_{1h}(r) &=& \int d^{3}\, \r' \int dM \phi(M)\,
 \rho_{\alpha}(M,r')\, \rho_{\alpha}(M,|\r'+\r|)\,, \nonumber\\
\bar{\rho}^{2} \xi_{2h}(r) &=& \int d^{3}\, \r' d^{3}\r''\,
 \left [\int dM_{1} \phi(M_{1}) 
\rho_{\alpha}(M_{1},r')\right] \nonumber\\
&& \times \left [\int dM_{2} \phi(M_{2})
 \rho_{\alpha}(M_{2},r'')\right] \nonumber\\
&& \times \xi_{hh}(|\r'-\r''+\r|,M_1,M_2)\,.
\end{eqnarray}
In the above equation, $\bar{\rho}$ is the mean mass density and
$\xi_{hh}(r,M_{1},M_{2})$ is the two-point correlation function of
halos of mass $M_1$ and mass $M_2$.  In k-space, using
$\tilde{\rho}_{\alpha}(M,k)$ to denote the Fourier transform of
$\rho_{\alpha}(M,r)$, where $\tilde{\rho}_{\alpha}(M,k) =
\int_{0}^{\infty}\, d^{3}\r \rho_{\alpha}(M,r)e^{-i\k\cdot \r}$ , we
can derive the mass power spectrum $P(k)$,
\begin{eqnarray}
P(k)&=&P_{1h}(k)+P_{2h}(k)\,,\nonumber\\
\bar{\rho}^{2} P_{1h}(k)&=&\int dM \phi(M)\, 
\tilde{\rho}^{2}_{\alpha}(M,k)\,, \nonumber\\
\bar{\rho}^{2} P_{2h}(k)&=&\int dM_{1} \phi(M_{1})\, 
\tilde{\rho}_{\alpha}\,(M_{1},k)\, 
\int dM_{2} \phi(M_{2})\,
\tilde{\rho}_{\alpha}(M_{2},k) \nonumber\\
&& \times P_{hh}(k,M_{1},M_{2})
\end{eqnarray}
Where $P_{hh}(k,M_{1},M_{2})$ is the Fourier transform of 
$\xi_{hh}(r,M_{1},M_{2})$, that is the cross power spectrum of the
halos of mass $M_1$ and $M_2$.  In the halo model, the mass power
spectrum or the mass correlation function is fully determined by
the halo mass function $\phi(M)$, the halo mass density profile
$\rho_{\alpha}(M,r)$ and the halo-halo power spectrum.

The halo mass function $\phi(M)$ is given by the Press-Schechter formula,
\begin{equation}
\phi(M) = \left [\frac{2} {\pi}\right]^{1/2} \left |\frac {d \ln
\sigma(M)} {d \ln M} \right|\, \frac{\bar{\rho}} {M}\, \nu
e^{-\nu^{2}/2}\,, \qquad \nu = \frac{\delta_{c}} {\sigma(M)}
\end{equation}
where $\delta_{c}$ is a parameter characterizing the linear
overdensity at the onset of gravitational collapse, which is $1.686$
in the case of a spherical collapse model in an Einstein-de Sitter
universe and depends weakly on cosmological models. The rms linear
mass fluctuation $\sigma(M)$ in spheres of radius R is related to the
linear power spectrum $P_{lin}(k)$
\begin{equation}
\sigma^{2}(M)=\int_{0}^{\infty} \frac{4 \pi k^{2} dk} {(2 \pi)^{3}}\,
P_{lin}(k) W^{2}(kR)
\end{equation}
\label{eq9}
where $W(x) = 3(\sin x-x\cos x)/ x^{3}$ is the Fourier transform of a
real-space top-hat window function.

We take the following form
\begin{equation}
\frac{\rho_{\alpha}(r)} {\bar{\rho}}= \frac{\bar{\delta}}
{(r/R_{s})^{p}(1+r/R_{s})^{3-p}}
\end{equation}
\label{eq10}
for the density profile of halos in CDM cosmological models.  In the
above equation, $\bar{\delta}$ is a dimensionless density amplitude,
and $R_{s}$ is a scale radius.  The seminal work of Navarro, 
Frenk, \& White (1996,1997; NFW) showed that the halos formed in 
CDM models can all be well described by  equation (10) with $p=1$.  
Subsequent work with higher numerical resolution has shown that the innermost
density profiles of the halos are steeper than the NFW form especially
for the halos of galactic mass (Fukushige \& Makino 1997; Moore et
al. 1998; Jing \& Suto 2000; Fukushige \& Makino 2001), and the
profiles are better described by  $p=1.5$. Defining the
radius of a halo, $R_{200}$, such that within it the mean matter density
is 200 times the mean density, the density profile is
characterized by the two parameters $p$ and the concentration
parameter $c=R_{200}/R_s$. The concentration parameter was found to
depend on the halo mass, and can be computed for $p=1$ 
semi-analytically (NFW). In this paper, we will use the following
fitting formula
\begin{eqnarray}
c(M)&=&7.0\left (\frac{M_{*}} {M}\right)^{\frac {1} {6}}\,, 
\qquad p=1\,, \nonumber\\
c(M)&=&3.5\left (\frac{M_{*}} {M}\right)^{\frac {1} {6}}\,, 
\qquad p=1.5,
\end{eqnarray}
for the LCDM model which fits well the concentration parameters
measured by Jing \& Suto (2000) for this model (see also Lokas
2001). A simple fit to the concentration parameters of SCDM halos in
NFW gives
\begin{equation}
c(M) =10 \left(\frac {M_{*}} {M}\right)^{0.09}\,, \quad p=1.0\,.
\end{equation}
\label{eq12}
and as for the LCDM, the concentration for $p=1.5$ is 0.5 times
of that for $p=1$.  The mass of a halo $M$ is related to $R_{200}$
simply by $M= \frac {800\pi} {3} R_{200}^3 \bar{\rho}$, and $M_{*}$ is defined by requiring $\sigma(M_{*})=\delta_{c}$.

The halo distribution is assumed to be linearly biased relative to the
underlying mass distribution, that is the power spectrum
$P_{hh}^{R}(k,M_{1},M_{2})$ is related to the mass power spectrum
$P_{mass}(k)$ 
\begin{equation}
P_{hh}^{R}(k,M_{1},M_{2})=b(M_{1}) b(M_{2}) P_{mass}(k)
\end{equation}
\label{eq13}
where $b(M)$ is the bias factor of halos for mass $M$. We will
use the analytical formula derived by Mo \& White (1996)
\begin{equation}
b(M)=1+(\nu^{2} - 1)/\delta_{c}\,, \qquad
\nu=\delta_{c}/\sigma(M)\,,
\end{equation}
for the bias factor.  Following Ma \& Fry (2000b), we will use the
linear density power spectrum $P_{lin}(k)$ for $P_{mass}(k)$, partly
for simplicity but more importantly because it can partly account for the
exclusion effect between halos. In \S 4, we test the accuracy of
this assumption.

Now, consider two cosmological models. The LCDM model is a
spatially flat model with matter density $\Omega_{m}=0.3$ and
cosmological constant $\Omega_{\Lambda}=0.7$. Its linear power
spectrum has the shape parameter $\Gamma = \Omega_m h =0.2$ and the
amplitude $\sigma_{8}=1$ (which is the rms mass density fluctuation in
a top-hat window with radius $8h^{-1}$Mpc). The SCDM model is a
spatially flat model but without a cosmological constant. Its linear
power spectrum has $\Gamma =0.5$ and $\sigma_{8}=0.6$.

\begin{figure}
%{\epsfxsize=10.1truecm \epsfysize=9.5truecm
%\epsfbox[95 230 535 680]{fig1.ps}}
\vskip-0.9cm
%\hskip-3.1cm
\centering
\mbox{\psfig{figure=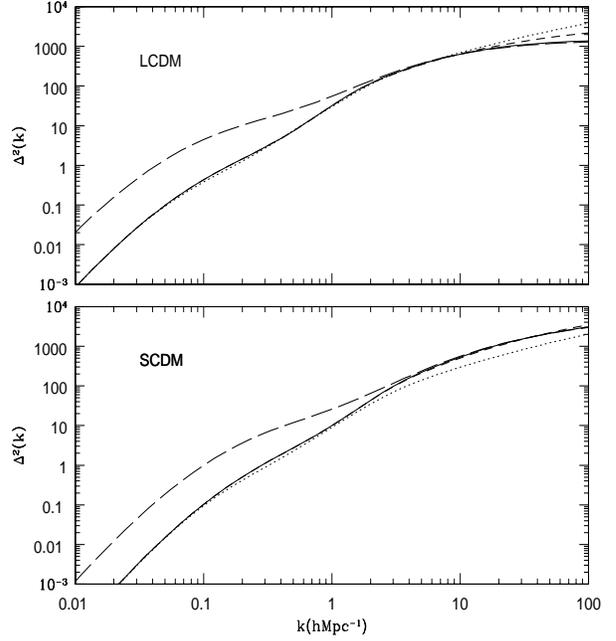,height=10.5cm,width=8.5cm}}
\vskip-1.0cm
\caption{The real-space power spectrum of dark matter in the LCDM and
SCDM models. The dotted line is the prediction based on the formula of
Peacock \& Dodds (1996), the solid (for $p=1.0$) and the short-dashed
(for $p=1.5$) lines are the predictions of the halo model with the
density profile cut at the radius $R_{200}$. The long-dashed line is
the prediction of the halo model for the density profile of p=1.0
without a cutoff in radius}
\end{figure}
\begin{figure}
\centering
\mbox{\psfig{figure=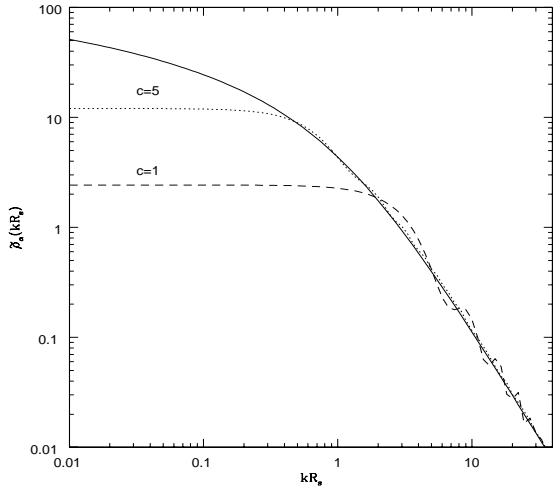,height=8.cm,width=8.cm}}
\caption{The Fourier transform of the density profile $\rho_{\alpha}(c,r)$. The dashed and the
dotted lines are for c=1 and c=5 respectively with the density profile cutoff at radius $R_{200}$.
 For comparison, the solid line is the fitting formula of Ma \& Fry
(2001b) in which the density profile extends to the infinity}
\end{figure}
Figure 1 shows a comparison between the real-space power spectrum
derived from the halo model and the fitting formula of Peacock \&
Dodds (1996; PD96) for the two cosmological models at redshift $z=0$.
First, we used the fitting formula of Ma \& Fry (2000b) for the
Fourier transform of the density profile
$\tilde{\rho}_{\alpha}(M,k)$. When obtaining the fitting formula they
apparently assumed that the density profile (equation 10) extends to
infinity. With their fitting formula, we obtained the mass power
spectrum for the LCDM model ( the long-dashed line in the upper panel) which
is in good agreement with the fitting formula of PD96 on small scales
($k \geq 2h^{-1}$Mpc), but has much more power on large scales.  The
reason is that the halo density profiles of equation (10) are assumed
to extend to infinity,  while there should be a radial cutoff in the
halo structure (Sheth, Hui, Diaferio \& Scooccimarro 2001). When we 
take $R_{200}$ as the radial cutoff for the
Fourier transform $\tilde{\rho}_{\alpha}(M,k) = \int_{0}^{R_{200}}
d^{3}\r \rho_{\alpha}(M,r) e^{-i\k \cdot \r}$, we get a mass power
spectrum which is in very good agreement with the fitting formula of
PD96 both on small and on large scales in the LCDM model. The
significance of the cutoff is illustrated in Figure 2 for $p=1$, where
we plot our computed $\tilde{\rho}_{\alpha}(M,k)$ against the
algebraic expression without a cutoff of Ma \& Fry (2000b) (the solid
line). The radial cutoff brings about a flat
$\tilde{\rho}_{\alpha}(M,k)$ at small $k$ (large scale), significantly smaller than without a radial cutoff for $k<
1/R_{200}$. As we have seen from Figure 1, the radial cutoff is
important to get the correct real-space power spectrum, and we will
adopt this cutoff throughout the paper (A cutoff was also imposed in
the work of Ma \& Fry 2001a,b; C.P. Ma, private communications). For
the SCDM model, the halo model prediction is 50 percent higher than
the fitting formula of PD96 on small scales (for $k \geq 1h^{-1}$Mpc).
A similar discrepancy is found for the LCDM model (the figure not plotted
in the paper) if we replace $\bar{\rho}$ (equation 10) with
$\rho_{crit}$ and take $R_{200}$ as the radius of a sphere of mean
interior density $200\rho_{crit}$, where $\rho_{crit}$ is the critical
density of the universe. Therefore, the inaccuracy of the halo model
prediction for the mass power spectrum is perhaps 50 to 100 percent
(in contrast to many recent studies which claimed that the halo model is
much more accurate). In a subsequent paper we will examine this
issue in a greater detail by comparing the halo model with the results
from high-resolution N-body simulations of $512^3$ particles.

\begin{figure}
\centering
\mbox{\psfig{figure=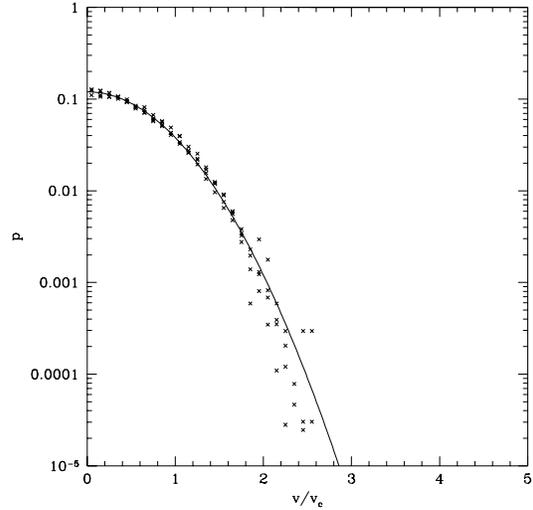,height=8.cm,width=8.cm}}
\caption{The 1-D velocity distribution of particles within virialized
halos in simulations (symbols) is well described by a Maxwellian distribution 
with $\sigma_{v}^{1D}/V_{200} = 1/1.53$ (the solid line)}
\end{figure}
%\begin{figure}
%{\epsfxsize=12.truecm \epsfysize=12.truecm
%\epsfbox[10 60 500 700]{fig3.ps}}
%\caption{ss3}
%\end{figure}
The mass power spectrum calculated in the halo model is quite
insensitive to the choice of $p$ for the innermost density
profile. The difference in $P(k)$ between $p=1$ and $p=1.5$ is rather
small for $k \le 10 \impc$, and we will consider only $p=1.5$  for the
calculation of the redshift-space power spectrum below. We also found
that the real-space power spectrum in the halo model is quite robust 
against
a reasonable change of the mass function $\phi(M)$ and the linear bias
factor $b(M)$. In contrast we will see that the red-shift power spectrum is
quite sensitive to the changes of these functions.

\section{The power spectrum in redshift space}
\begin{figure}
\centering
\mbox{\psfig{figure=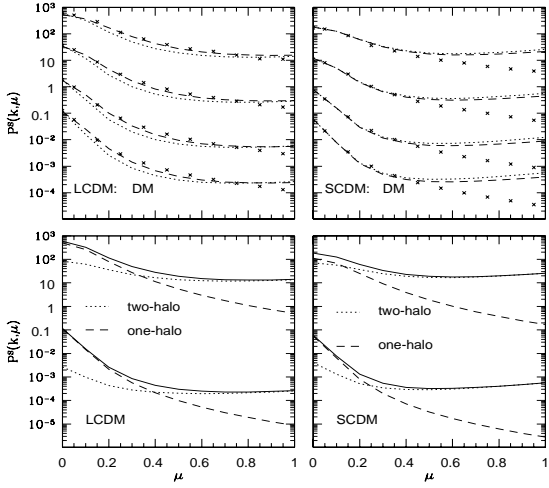,height=8.cm,width=8.cm}}
\vskip-0.5cm
\caption{The redshift power spectrum of dark matter (dot lines in the
upper panels) and contributions from the one-halo and the two-halo 
terms (the lower panels). The dashed lines in the upper two panels 
are the results after accounting for the finite size of the simulation 
box. The crosses are the simulation results of JB2001. In the upper 
panel, from top to bottom, the values of $k$ are 1.1, 1.7, 2.7, and 
3.4 hMpc$^{-1}$ respectively, and the values
$P^{s}(k,\mu)$ are multiplied by $1$, $10^{-1}$, $10^{-2}$, and
$10^{-3}$ respectively for clarity. For the halo model, the
Press-Schechter formula, the formula of Mo \& White (1996), and 
Kaiser's formula are used for the mass function of halos, the halo
bias and the redshift distortion of halos. In the lower panel, the
one-halo and two-halo contributions are drawn by  dotted and dashed
lines. Their sums are the solid lines. The wavenumber is
1.1 and 3.4 hMpc$^{-1}$ for the upper and lower lines and 
the $P^{S}(k,\mu)$ values  have been multiplied by 1 and $10^{-3}$ 
respectively for clarity.}
\end{figure}
The dark matter halo is nearly virialized inside $R_{200}$. The
velocity distribution of the dark matter within the halo should be
approximately Maxwellian distributed with a one-dimensional
velocity dispersion $\sigma_{v}^{1D}$ (Sheth 1996). We have tested
this assumption with the N-body simulations of Jing \& Suto
(1998), and confirmed that the assumption is valid (see
Figure 3). Furthermore, we found that
\begin{equation}
\sigma^{1D}_{v} / V_{200} \approx 1/1.53
\end{equation}
\label{eq15} slightly smaller than $1/\sqrt{2}$, where $V_{200} =
\sqrt{GM/R_{200}}$  is the circular velocity at the viral radius
$R_{200}$, and $M$ is the mass of the halo within $R_{200}$. The
virial motion in the halo elongates the density distribution of
the halo along the line-of-sight in redshift space . The 
density distribution of a halo in redshift space 
$\rho_{\alpha}^{S}(r_{\pi},r_{\sigma})$ is a convolution of the 
real-space density profile with the velocity distribution along 
the line-of-sight,
\begin{equation}
\rho_{\alpha}^{S}(r_{\pi},r_{\sigma}) = \int \rho_{\alpha}\, 
(r_{\pi}-v_{z}/H,r_{\sigma})\, f(v_{z})dv_{z}\,,
\end{equation}
\label{eq16}
where $r_{\pi}$ and $r_{\sigma}$ are the distances parallel and
perpendicular to the line-of-sight to the halo center, $H$ is 
the Hubble constant, and $f(v_{z})$ is the velocity distribution 
along the line-of-sight.  The redshift space power
spectrum $P^{S}(k,u)$ can then be written as,
\begin{eqnarray}
P^{S}(k,\mu)&=& P_{1h}(k,\mu)+P_{2h}(k,\mu) \nonumber\\
P_{1h}^{S}(k,\mu) &=& \int dM \phi(M)\,
\tilde{\rho}^{2}_{\alpha}(M,k)\,
e^{-k^{2}\mu^{2}\, \sigma_{v}^{{1D}^{2}(M)}}\,, \nonumber\\
P_{2h}^{S}(k,\mu) &=& \int dM_{1} \phi(M_{1})\,
\tilde{\rho}_{\alpha}\,
(M_{1},k) e^{-k^{2}\mu^{2}\,
 \sigma_{v}^{{1D}^{2}}(M_1)/2} \nonumber\\
& \times & \int dM_{2} \phi(M_{2})\, 
\tilde{\rho}_{\alpha}(M_{2},k)\, e^{-k^{2}\mu^{2}\,
 \sigma_{v}^{{1D}^{2}}(M_2)/2} \nonumber\\
& \times & P^{S}_{hh}(k,\mu,M_{1},M_{2})\,.
\end{eqnarray}
Where $P^{S}_{hh}(k,\mu,M_{1},M_{2})$ is the redshift power spectrum
for two dark matter halos with mass $M_{1}$ and $M_{2}$. Under the
assumption that every halo is moving as a whole according to 
linear theory (Kaiser 1987) , we can derive the redshift power
spectrum for halos,
\begin{eqnarray}
P^{S}_{hh}(k,\mu,M_{1},M_{2})&=& 
\left[1+\Omega^{0.6}\mu^{2}/b(M_{1})\right]
P_{hh}^{R}(k,M_{1},M_{2})\nonumber\\
& \times & \left[1+\Omega^{0.6}\mu^{2}/b(M_{2})\right]\,.
\end{eqnarray}
We will show that the formula is a good description on large scales,
but on small scales it gives too much power at large $\mu$.  A modified
formula will be given in the next section.
\begin{figure}
\centering
\mbox{\psfig{figure=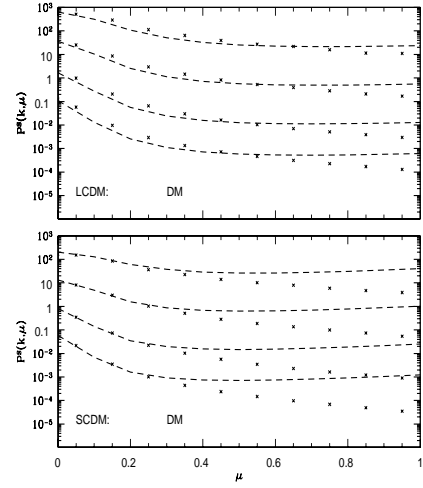,height=8.cm,width=6.cm}}
\vskip-0.5cm
\caption{The same as the upper panels of Figure 4 but with the modified
formulae for the halo mass function and the linear bias factor}
\end{figure}

In Figure 4 we plot the redshift power spectrum for dark matter
based on this halo model, and compare it to the
simulation results of JB2001 for the two cosmological models. From
the dot line in the top panels, we see that the difference in the
redshift power spectrum between the halo model and the simulations is small on
large scales (i.e small $k$). But on small scales the halo model
gives significantly more power at larger $\mu$ than the simulation
data, especially in the SCDM model. It is also noted that the
finite box size effect should be taken into account. In Figure 4 
we also plot $P^{S}(k,\mu)$( the dashed lines in the upper two panels) 
calculated using a cut-off of the power spectrum for a
scale smaller than the box-size (100$h^{-1}$Mpc in our simulation). 
The value of $P^{S}(k,\mu)$ at larger $\mu$ are lower in the case, but
the changes are quite small. The difference is so
significant that we have to check all the assumptions used in our
model: the halo mass function; the density profile of halos; and
the halo-halo redshift power spectrum. Before doing that, it would
be helpful first to check which term, the one-halo term or two
halo term, is the main contribution to the discrepancy. From the
lower panels of Figure 4, it can be easily found that the two-halo
term dominates the redshift power spectrum at large $\mu$ even for
small scales ($k$ about a few $\impc$), in contrast to the
real-space power spectrum (Ma \& Fry 2000b). The reason for the 
difference is that the 
strong real-space clustering power from the one-halo contribution 
is suppressed by its internal motion, while the infall (merging) 
of halos to each other enhances  the $P^S(k,\mu)$. We also find 
that the small halos contribute significantly to the two-halo term.  
As for the real-space power spectrum, we find that the density profile 
of dark matter has little effect on $P^S(k,\mu)$ for 
$k \leq$ 4hMpc$^{-1}$ (the upper limits of k in our paper) for 
$1\le p \le 1.5$. In the following section, we will check the 
validity of the assumptions about the mass function and the 
halo-halo correlation.

\section{Modifications of the assumptions about the halo distributions}
\begin{figure}
\centering
\vskip-0.4cm
\mbox{\psfig{figure=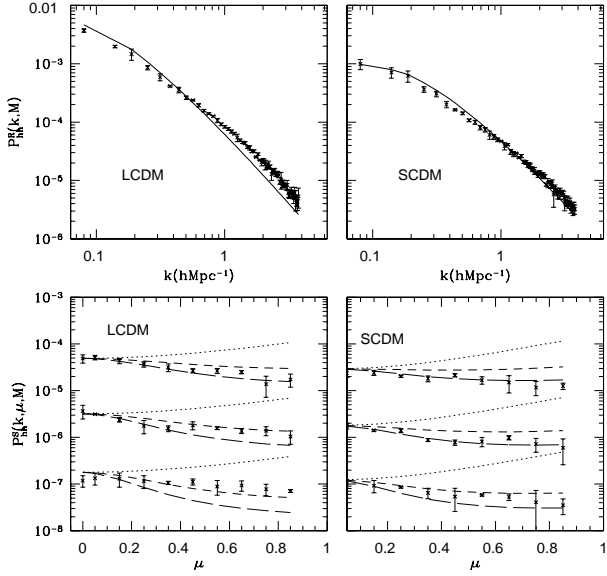,height=9.2cm,width=8.5cm}}
\vskip-0.4cm
\caption{Upper panels -- The real-space power spectrum of halos with
mass $M=1.75 \times 10^{11}h^{-1}M_{\odot}$ for the LCDM and $M=6.0
\times 10^{11}h^{-1}M_{\odot}$ for the SCDM model measured from the N-body
simulations (the symbols) compared with the formula of Eq.(13) (the solid
lines). Lower panels -- the redshift space power spectrum of halos in the
simulations. The values of k are 1.4, 1.7 and 2.1 hMpc$^{-1}$ respectively.
 The dotted , the dashed and the long-dashed lines are the
predictions of equation (20) with $\sigma=0$ , 200, and $300\kms$ and
the nonlinear real-space power spectrum of halos. The 
$P_{hh}^{S}(k,\mu,M)$ values have been multiplied by 1,
$10^{-1}$,$10^{-2}$ respectively for clarity}
\end{figure}

It has been shown by Jing (1998) and Lee \& Shandarin (1998)
that the Press-Schechter formula, equation (8) and equation (14)
for the halo bias are not good descriptions 
for small halos in high resolution simulations.  Actually the
simulations produce fewer halos than the P-S formula in an intermediate
mass range $0.2 \leq M/M_{*} \leq 1$, but more halos both for
smaller and larger masses. Motivated by these 
findings, \cite{ST1999} have found the fitting formula for the mass 
function and the halo bias parameter (see also Sheth, Mo \& Tormen 2001),
\begin{eqnarray}
\phi(M)&=&\frac{\bar{\rho}} {M^{2}} \frac{d \log \nu'} 
{d \log M}\, A \left[1+ \frac{1} {a\nu'^{n}}\, 
(\frac{a\nu'} {2})^{1/2}\,
 \frac{e^{-a\nu'/2}}
{\sqrt{\pi}}\right]\,, \nonumber\\
b(M)&=&1 + \frac{a\nu' - 1} {\delta_{c}}
\, + \frac{2n/\delta_{c}} {1+(a\nu')^{n}},
\end{eqnarray}
where $a = 0.707$ , $n = 0.3$, $A=0.32$ and
$\nu'=(\delta_{c}/\sigma(M))^{2}$. We now take these formula and recalculate
the redshift-space power spectrum for dark matter. The results are
shown in Figure 5, where we can find that at small $\mu$ the agreement
between the model and the simulation is improved very moderately, but
not at larger $\mu$. The results can
be explained as follows: the modified formulae for the halo mass
function and the halo bias give a higher number density and
a higher bias factor for small halos (which have mass less than
$10^{11}M_{\odot}$) than the P-S and Mo \& White
formulae used in the last section, both of which boost the
two-halo term of the real-space power spectrum. While the linear
redshift compression due to the Kaiser effect (which increases the
redshift power spectrum) is reduced by a larger bias, the net
effect of using the accurate fitting formula enhances the redshift
power spectrum at large $\mu$ and at large $k$, thus spoiling the agreement
of the halo model with the simulation .

\begin{figure}
{\epsfxsize=6.5truecm \epsfysize=9.2truecm
\epsfbox[50 300 420 705]{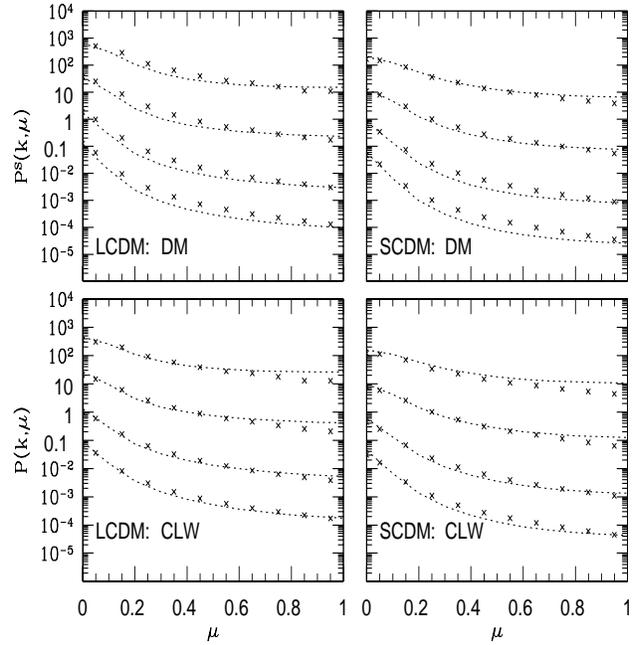}}
%\vskip-0.9cm
%\centering
%\mbox{\psfig{figure=fig7.ps,height=8.4cm,width=8.cm}}
\caption{Upper panels -- the same as Figure 5, but with the nonlinear
motions of halos included in the halo model. The parameter $\sigma$ is
$200\kms$ for the LCDM and $300\kms$ for the SCDM model. Lower panels
-- the same as the upper panels but for galaxies described by the
cluster weighted model}
\end{figure}

Considering carefully all the assumptions in the halo model which
may cause the discrepancy with the
simulation results, we find only one assumption still needs to be
checked, i.e. the redshift-space power spectrum of the halos. The
halos were assumed to be linearly biased relative to the {\it
linearly evolved matter distribution}, with the velocity distortion
fully described by Kaiser's formula. During the non-linear
evolution both assumptions about the
redshift-space power spectrum of the halos
may break down. Here we check the
assumptions with the help of the N-body simulations of the box
size $100\mpc$ of Jing \& Suto (1998). A comparison of the real
space power spectrum of the halos in the simulations with equation
(13) for halo masses around $M=3 \times 10^{11}M_{\odot}$ (the halos
around this mass are most important for the two-halo term) are
shown in the upper panel of Figure 6, where the fitting formula of
Jing (1998) is used for the linear bias factors $b(M)$. Overall,
equation (13) gives a reasonable approximation for the real space
power spectrum of halos in the simulations of both CDM models,
with the difference less than a factor of 2. More precisely, for
the small scales $k\sim 2hMpc^{-1}$ 
the simulation results are about a factor of 2 higher
than the adopted formula in the LCDM, but the difference between
equation (13) and the simulations is much smaller in the SCDM
model. From these comparisons, we are sure that the differences in
the real-space power spectrum between the simulations and the
adopted simple formula cannot fully account for the discrepancy found
for the redshift space power spectrum in Figure 4.
\begin{figure}
\centering
\vskip-0.4cm
\mbox{\psfig{figure=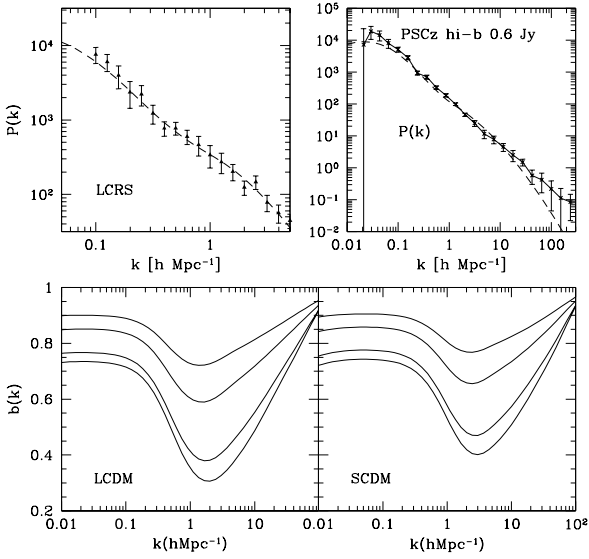,height=8.8cm,width=8.2cm}}
\caption{Upper panels --- the real-space power spectrum of
galaxies. The left panel is for the optical galaxies of the Las
Campanas Redshift Survey (the data from Jing \& B\"orner 2001) and the
right panel is for the IRAS galaxies of the PSCz survey ( the data
from Hamilton \& Tegmark 2000). The dashed lines are the predictions
of the LCDM model with $\alpha = 0.08$ in the left panel and 0.25 in
the right panel. Lower panels --- the scale dependent bias $b(k)$
predicted by the cluster weighted bias model for the two CDM
models. From top to bottom, the value of $\alpha$ are 0.08, 0.13,
0.25, 0.32}
\end{figure}

The final possible cause for the discrepancy is that the redshift
distortion of halos is not precisely described by Kaiser's
formula. Now we check the power spectrum of halos in redshift
space using the simulation data of Jing \& Suto (1998). Some typical
examples (the symbols) are shown in the lower panels of Figure 6.  For
comparison, the linear prediction of equation (18) is plotted (the
dotted lines), where the real-space power spectrum from the simulations
is used for $P^R(k,M_1,M_2)$. The linear prediction is much higher
than the simulation results for larger
values of  $\mu$. Obviously, the
redshift distortion of halos is not well described by Kaiser's
formula. The result is not surprising, as we have noted that the peculiar
velocity of massive halos deviates typically from the 
prediction of linear theory
by $\sim 30\%$ (Colberg et al. 1998). We parameterize the
non-linear peculiar velocity effect by a Lorentz damping factor,
\begin{eqnarray}
P^{S}(k,\mu,M_1,M_2)&=&\frac {P_{hh}^{R}(k,M_1,M_2)}
{1+(k\mu\sigma_{h})^{2}/2}
\left[1+\Omega^{0.6}\mu^{2}/b(M_{1})\right]
\nonumber\\
& \times & \left[1+\Omega^{0.6}\mu^{2}/b(M_{2})\right]\,.
\end{eqnarray}
where $\sigma_{h}$ is a parameter characterizing the
pairwise velocity dispersion of halos. The parameter $\sigma_{h}$ may
be a function of halo mass $M_{1}$, $M_{2}$ and $k$, but Sheth \&
Diaferio (2001) show that $\sigma_{h}$ is insensitive to the halo mass.
So here we treat it as a constant. This is a valid approximation, 
as shown in Figure 6. The dashed lines in the lower left panel and the long-dashed lines
in the lower right panel agree quite well with the simulation
results. The corresponding $\sigma_{h}$ values are $200\kms$ for the
LCDM model and $300\kms$ for the SCDM model, qualitatively
consistent with the findings of Colberg et al. (1998).  In the
upper panel of Figure 7 we present the redshift power spectrum for
dark matter using the three modifications discussed above: the modified
Press-Schechter function, the modified halo bias (including the
non-linear bias from the simulation), and the non-linear peculiar
velocity of halos.  The result is in very good agreement with the
simulation result of $P^{S}(k,\mu)$ at all scales (with a
difference less than a factor of 2).

 In a recent paper, Padilla \& Baugh (2002) studied the power
spectrum of galaxy clusters in redshift space using large N-body
simulations. They found that there is no damping of the power spectrum 
at high wavenumber in redshift space. It is difficult to make direct
comparison between our results and theirs because the halos they used
in their analysis are only massive ones, which are separated by large
distances and the nonlinear effects may be insignificant.

\section{The power spectrum of galaxies}
\begin{figure}
\centering
\vskip-0.4cm
\mbox{\psfig{figure=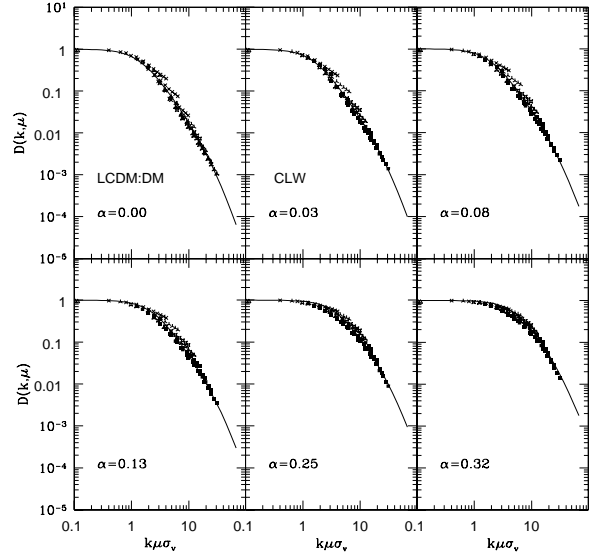,height=8.8cm,width=8.2cm}}
\caption{The damping factor $D(k,\mu,\alpha)$ for the LCDM model. The
solid lines are our fitting formula (Equation 23) with $\sigma_{v}$
fixed to $1100 \kms$. The values of $\alpha$ are inserted in each plot}
\end{figure}

This halo model can be readily extended to the real-space and
redshift-space power spectra for galaxies if we assume a model for the
number of galaxies $N$ per halo of mass $M$ (or the halo occupation
number). It was shown by Jing, Mo \& B\"orner (1998) that the
LCDM model with a power-law halo occupation number model,
\begin{equation}
N/M \propto (M/M_{0})^{-\alpha}
\end{equation}
\label{eq21} can accurately explain the observed two-point
correlation function and the pairwise velocity dispersion of the
galaxies in the Las Campanas Redshift Survey, where $\alpha = 0$ for
$M \leq M_{0}$, and $\alpha = 0.08$ for $M \geq M_{0}$ and $M_{0}$ is
$6\times 10^{11}h^{-1}\msun$. Since $\alpha$ is positive for $M \geq
M_{0}$, i.e. the number of galaxies per unit dark matter mass within a
massive halo decreases with the halo mass, JMB98 called this model the
cluster-(under)weighted model.  They also noted that this
parameterization is consistent with the observed trend of Carlberg et
al. (1996) for the CNOC clusters.  In upper left panel of Figure 8, we
compare our result with the real space power spectrum of the Las
Campanas Redshift Survey (LCRS).  The dashed line is our result for LCDM 
model and the value of $\alpha$ is $0.08$. There is good agreement on all
the scales.
\begin{table}
 \begin{center}
  \begin{tabular}{lcccccc}
   Model & $A_{1}$ & $\gamma_{1}$ & $\eta$ & $A_{2}$ & $\gamma_{2}$
 \\
    & & & & & & \\
   SCDM  & 4.43 & 0.84 & 5.44$\times10^{-3}$ & 8.27 & 0.90  \\
   LCDM  & 8.41 & 1.05 & 6.75$\times10^{-4}$ & 9.22 & 0.84  \\
     \end{tabular}
 \end{center}
\caption{The fitting values of the coefficients in Equation 23}
\end{table}

It has been recently shown that this model can also successfully
describe the clustering of the IRAS galaxies in the PSCz catalog when
a higher $\alpha \approx 0.25$ is assumed (Jing, B\"orner, \& Suto
2001; Scoccimarro \& Sheth 2001). The upper right panel of Figure 8
shows a comparison of the halo model with the real space power
spectrum measured by Hamilton \& Tegmark (2000) for the survey. In the
figure a value of $\alpha = 0.25$ is assumed, and the model power
spectrum is in good agreement with the observed result.  Although the
spatial bias of the IRAS galaxies is scale-dependent compared with the
predictions of CDM models as Hamilton \& Tegmark (2000) emphasized,
such non-trivial properties of the spatial bias are actually predicted
by the cluster weighted halo model (see also Jing et al. 1998).  The
scale-dependent bias $b(k)=(P(k,\alpha)/P(k,\alpha=0))^{1/2}$ is
plotted in the lower panels of Figure 8 for several choices of
$\alpha$ in the two CDM models. The bias factor has a nontrivial
dependence on scale for $k>0.1\impc$. It could be regarded as
scale-invariant (or a linear bias) only on larger scales. Since many
observation, like the dynamical measurement of
$\beta$ based on measuring the peculiar velocity and spatial
distribution of galaxies, rely on the assumption of a linear bias, it
is very important to note that such observations must be carried out
on sufficiently large scales (also see Seljak 2001)

We also calculate the redshift space power spectrum of
galaxies for $\alpha$ varying from 0.03 to 0.32. The analytical result 
(the lower panels of Figure 7) agrees pretty well with the simulation 
result of JB2001 for the same cluster-weighted model, where 
$\alpha = 0.08$ is taken. Following JB2001, we consider the damping factor
\begin{equation}
D(k,\mu,\alpha) = \frac{P^{S}(k,\mu,\alpha)} {P^{R}(k,\alpha)
(1+\beta\mu^{2})^{2}}\,.
\end{equation}
\label{eq22}                     
which describes the non-linear motion effect on the redshift power spectrum. 
We have fitted the scaled damping factor with the following form
\begin{equation}
D(k,\mu,\alpha) = \frac{1} {1+\frac{1} {2}f(\alpha)\,
(k\mu\sigma_{v})^{2}
\, +\eta g(\alpha)k\mu\sigma_{v})^{4}}\,.
\end{equation}
which takes into account the fact that the damping function is
approximately a scaling function of $k\mu\sigma_{v}$. $f(\alpha)$
and $g(\alpha)$ have the forms $e^{-A_1\alpha^{\gamma_1}}$and
$e^{-A_2\alpha^{\gamma_2}}$ respectively. The fitting values for the
parameters are given in Table~1 for the two cosmological models. The
effective pairwise velocity dispersion is $f^{1/2}(\alpha)\sigma_v$,
(where $\sigma_v$ is the velocity dispersion corresponding to $\alpha
= 0$ ,ie  $f(0) = 1$), which describes  how the velocity dispersion
changes with the parameter $\alpha$.  Also note that $f(\alpha)$
depends on the cosmological model and the power spectrum. In our model
for dark matter, the effective pairwise velocity dispersion decreases
by $40\%$ when $\sigma_{8}$ changes from 1 to 0.7. The formula can
also be used to compare the theoretical models with the statistic in 
large redshift surveys of galaxies.

This CLW model has been shown to be qualitatively consistent with the
semi-analytical models of galaxy formation which take into account
star formation (Sheth et al. 2001; White 2001). This type of model has
been applied to CDM models to predict various statistics for the
galaxy clustering(Jing, Mo \& B\"orner 1998; Peacock \& Smith 2000;
Seljak 2000; Scoccimarro et al. 2000; Berlind \& Weinberg 2001).  The
good agreement shown above means that the halo model can be
successfully applied to predict the power spectrum of various galaxies
once their halo occupation number model is specified. Thus insight on
 the galaxy formation can be obtained from measuring the clustering 
of different populations of galaxies.

\section{Discussion and Conclusions}
\begin{figure}
\centering
\vskip-0.7cm
\mbox{\psfig{figure=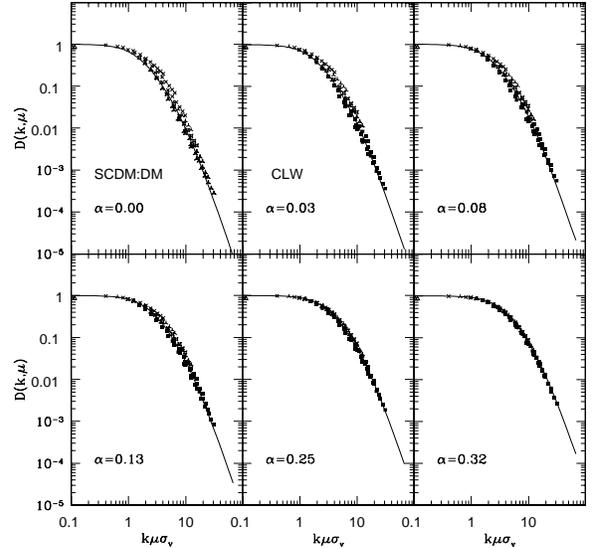,height=8.8cm,width=8.2cm}}
\caption{The same as Figure 9, but for SCDM model.}
\end{figure}

In this paper we present an analytical model for the non-linear
redshift-space power spectrum of dark matter and of galaxies based
on a halo prescription. The model has three important ingredients:
the halo mass function, the mass density profile of halos, and the
halo-halo redshift space power spectrum. The predicted redshift
power spectrum is found to be insensitive to the details of the
halo density profiles: the NFW density profile and a steeper inner
density profile (Jing \& Suto 2000) yield an indistinguishable
redshift power spectrum. When we use, as many others have done
(for the real space correlation function of dark
matter and for the radial averaged redshift power spectrum) the
Press-Schechter formula for the halo mass function, the Mo \&
White formula for the halo bias, and Kaiser's formula for the
redshift distortion of the halo spatial distribution, we find that
the predicted redshift-space power spectrum for dark matter is too
high at large $\mu$ to be consistent with the high-resolution simulation
results of JB2001. The reason why the halo model works well for 
the real space clustering but not for the redshift-space power 
spectrum is that in the latter case the result is dominated by 
the two-halo term on non-linear scales.

We have checked carefully which assumptions have caused the
discrepancy between the halo model prediction and the
simulations. First the fitting formulae of the halo mass function and
the halo bias from numerical simulations (Sheth \& Torman 1999) are
used in the replacement of the analytical formulae based on the
Press-Schechter formalism.  This deteriorates the agreement between
the halo model prediction and the simulation results , because there
are more small halos and the spatial clustering of small halos is
stronger in the modified formulae. We also found that the bias of the
halo distribution is nearly linear relative to the linear density
field, so the non-linearity of the bias could not be the main
contribution to the discrepancy. Instead we found that the non-linear
motions of the halos, which were neglected in previous studies, are
the main cause. Once we take this effect into account the redshift
power spectrum based on the halo model agrees very well with the
simulation results. Furthermore, the redshift space power spectrum can
be precisely predicted if the halo occupation number model, e.g. the
cluster weighted model, is given for the galaxies.

Our results show for the first time that the two-halo term can
dominate some statistics in the redshift space even at small
scales. Therefore a halo model based on the density profiles of halos
and on the redshift distribution of halos predicted by the {\it linear
theory} may become inaccurate for some statistics of redshift
clustering. For the time being, an accurate model (analytical or
fitting) for the redshift power spectrum of the halos, which includes
the effects of the nonlinear bias and the non-linear motions, is
needed for the prediction of the redshift space power spectra of
galaxies and dark matter.

In combination with the cluster weighted bias model, we show that a
non-trivial scale-dependent bias is generally expected for galaxies in
CDM models. The bias could be regarded linear only on the scales at
the wavelength larger than $60\mpc$ (also see Seljak 2001). The bias is different for
different populations of galaxies as observed. The currently favored
LCDM model can well explain the observed features of the spatial bias
reported for optical galaxies and for IRAS galaxies recently, if the
cluster weighted bias model is applied.

\section{Acknowledgments}
The work is supported in part by the One-Hundred-Talent Program, 
by NKBRSF(G19990754),  by NSFC(No.10043004), and by SFB375. We thank
an anonymous referee for helpful comments.

\end{document}